\documentclass[aps,prd,amsmath,two column,amssymb,showpacs]{revtex4}
\usepackage{amssymb}
\usepackage{txfonts}
\usepackage{mathbbol}
\usepackage{amsfonts}
\usepackage{mathrsfs}
\usepackage{epsfig,bm,dcolumn}
\usepackage{graphicx}
\usepackage{color}
\usepackage{amsmath}
\usepackage{dcolumn}
\usepackage{overpic}
\usepackage{slashed}

\begin{document}
\title{Chiral symmetry breaking in a semi-localized magnetic field}

\author{Gaoqing Cao$^{1,2}$}
\affiliation{1 School of Physics and Astronomy, Sun Yat-Sen University, Guangzhou 510275, China\\
	2 Department of Physics and Center for Particle Physics and Field Theory, Fudan University, Shanghai 200433, China.}
\date{\today}
\begin{abstract}
	In this work, I'm going to explore the pattern of chiral symmetry breaking and restoration in a solvable magnetic distribution within Nambu--Jona-Lasinio model. The special semi-localized static magnetic field can roughly simulate the realistic situation in peripheral heavy ion collisions, thus the study is important for the dynamical evolution of quark matter. I find that the magnetic field dependent contribution from discrete spectra usually dominates over that from continuum ones and chiral symmetry breaking is locally catalyzed by both the magnitude and scale of the magnetic field.
\end{abstract}

\maketitle
\section{Introduction}
Recently, because of both extremely strong electromagnetic (EM) field generated in peripheral relativistic heavy-ion collisions~(HICs), such as in Relativistic Heavy Ion Collider (RHIC) at BNL and Large Hadron Collider (LHC) at CERN~\cite{Skokov:2009qp,Voronyuk:2011jd,Bzdak:2011yy,Deng:2012pc}, and unexpected inverse magnetic catalysis effect (IMCE) at finite temperature from lattice quantum chromodynamics (LQCD) simulations~\cite{Bali:2011qj,Bali:2012zg,Bruckmann:2013oba,Endrodi:2015oba}, a lot of efforts were devoted to explaining or exploring the thermodynamic properties of strong coupling systems in the presence of constant magnetic field~\cite{Fukushima:2012kc,Chao:2013qpa,Feng:2014bpa,Cao:2014uva,Mueller:2015fka,Guo:2015nsa,Cao:2015xja,Cao:2015cka,Bonati:2016kxj}, see review Ref.~\cite{Miransky:2015ava}. Besides, due to the successful realization of chiral magnetic effect (CME) in the condensed matter system ZrTe5~\cite{Li:2014bha}, chiral anomaly phenomena~\cite{Hattori:2016njk,Qiu:2016hzd} and the related phenomenology in hydrodynamics~\cite{Huang:2015fqj,Jiang:2015cva,Jiang:2016wve} become even hotter topics which further push the efforts to look for the CME signal in QCD system, see reviews Ref.~\cite{Liao:2014ava,Kharzeev:2015kna,Huang:2015oca,Kharzeev:2015znc}. It is very interesting to notice that magnetic field usually brings us a lot of surprises due to the specific quantum effects. 

\begin{figure}[!htb]
	\begin{center}
		\includegraphics[width=8cm]{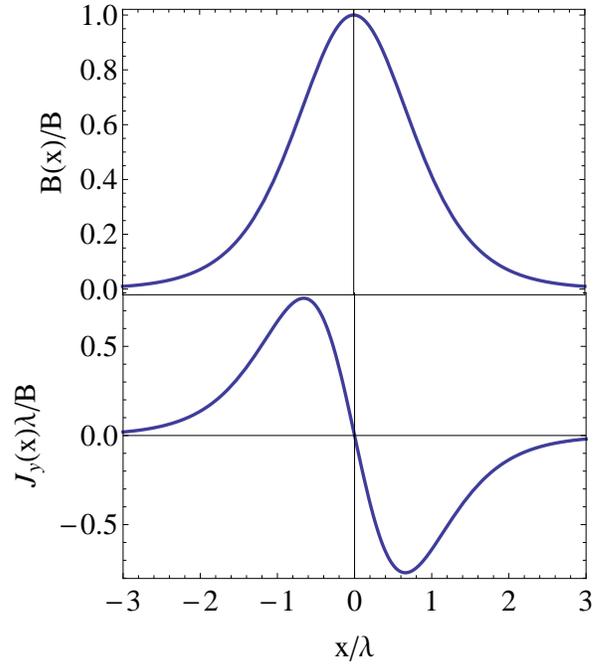}
		\caption{The distributions of the magnetic field $B(x)=B~{\rm sech}^2(x/\lambda)$ along $z$ direction and the corresponding current $J_y=B'(x)$ along $y$ direction in the region $(-3\lambda,3\lambda)$.\label{BJx}}
	\end{center}
\end{figure}

One thing that one should keep in mind about the HICs is that the magnetic field produced there is actually inhomogeneous in the fireball. Thus, it is very important to explore how the free energy and chiral symmetry will be affected by such a kind of magnetic field, which is the main goal of this work. Surely, the distribution of the magnetic field in the fireball is quite complicated due to initial charge fluctuation and later expansion of fireball, but the main feature can be captured by the distribution between two long straight electric currents with opposite directions~\cite{Deng:2012pc}. In order to derive an exact fermion propagator for later use, we choose an ideal semi-localized distribution, that is $\vec{B}(x)=B~{\rm sech}^2(x/\lambda)~\hat{z}$~\cite{Cangemi:1995ee,Dunne:1997kw}, which is illuminated in Fig.\ref{BJx}. As we can see, the corresponding electric current distribution is mainly composed of two peaks along opposite directions which is just like the case in peripheral heavy ion collisions. Previously, the contribution of fermions to free energy in such a magnetic field were studied in detail in both $2+1$~\cite{Cangemi:1995ee} and $3+1$ dimensions~\cite{Dunne:1997kw}, but the simple results (refer to Eq.(\ref{omegaB2})) should be treated cautiously since the quadratic terms of $B$ were not dropped completely. Quite recently, the Schwinger mechanism was checked and pair production was found to be enhanced in such a magnetic field together with the presence of a parallel electric field~\cite{Copinger:2016llk}.

In this work, I only focus on the pure magnetic field case for simplicity and the paper is arranged as the following: In Sec.\ref{NJL}, I develop the main formalism for a semi-localized magnetic field within Nambu--Jona-Lasinio (NJL) model, where Sec.\ref{potential} is devoted to calculating the thermodynamic potential of fermion systems preliminarily with the assumption of constant mass gap and Sec.\ref{WB} is devoted to exploring the pattern of chiral symmetry breaking in the weak magnetic field approximation. The main numerical results for both the constant and semi-localized mass gap ansatzs are given in Sec.\ref{results}. Finally, we briefly summarize in Sec.\ref{conclusions}.

\section{Nambu--Jona-Lasinio model with semi-localized magnetic field}\label{NJL}
In order to study the effect of semi-localized magnetic field to the chiral symmetry breaking and restoration in QCD systems, we adopt the effective Nambu--Jona-Lasinio model~\cite{Nambu:1961tp,Nambu:1961fr,Klevansky:1992qe} which has an approximate chiral symmetry as the basic QCD theory. Taking the magnetic field $B(x)$ and baryon chemical potential $\mu$ into account, the Lagrangian is give by
\begin{eqnarray}
{\cal L}=\bar\psi\left(i\slashed{D}-m_0+{\mu}\gamma^0\right)\psi
+{G}\left[\left(\bar\psi\psi\right)^2+\left(\bar\psi i\gamma^5{\boldsymbol \tau}\psi\right)^2\right],
\end{eqnarray}
where $\psi=\left(u,d\right)^T$ is the two-flavor quark field, ${\boldsymbol \tau}$ are pauli matrices in flavor space, $m_0$ is the current quark mass, and $G$ is the coupling constant with a dimension $[GeV^{-2}]$. Here, $D_\mu=\partial_\mu+iq A_\mu$ is the covariant derivative in flavor space with electric charges $q_u=2e/3$ and $q_d=-e/3$ for $u$ and $d$ quarks, and the static magnetic field is chosen to be along $z$ direction but vary along $x$ direction with the corresponding vector potential given by $A_\mu=(0,0,-B\lambda\tanh(x/\lambda),0)$~\cite{Cangemi:1995ee}.

In order to explore the ground state of the system, we introduce four auxiliary fields $\sigma=-2G\bar\psi\psi$ and ${\boldsymbol \pi}=-2G\bar\psi i\gamma^5{\boldsymbol \tau}\psi$, then the Lagrangian density becomes
\begin{eqnarray}
{\cal L} &=& \bar\psi\bigg[i{\slashed D}-m_0-\sigma-i\gamma^5\left(\tau_3\pi_0+\tau_\pm\pi_\pm\right)+{\mu}\gamma^0\bigg]\psi\nonumber\\
&&-{\sigma^2+\pi_0^2+\pi_\mp\pi_\pm\over 4G},
\end{eqnarray}
where $\pi_\pm$ are the physical fields which are related to the auxiliary fields as $\pi_\pm=\left(\pi_1\mp i\pi_2\right)/\sqrt 2$, and $\tau_\pm=\left(\tau_1\pm i\tau_2\right)/\sqrt 2$ are the raising and lowering operators in flavor space, respectively. The order parameters for the spontaneous breaking of $SU_L(3)\times SU_R(3)$ chiral symmetry are the expectation values of the collective fields $\langle\sigma\rangle, \langle\pi_0\rangle$ and $\langle\pi_\pm\rangle$. There should be no pion superfluid for vanishing isospin chemical potential, that is, $\langle\pi_\pm\rangle=0$ in the recent case. But there might exist stable $\pi^0$ domain wall due to the coupling term $\mu \bf{B\cdot\nabla}\pi^0(x)$ when $B$ exceeds the critical value $(0.255 {\rm GeV})^2$ in nuclear matter~\cite{Son:2007ny}. Though, we can show the possibility of $\pi^0$ domain wall in NJL model by expanding over small $\pi^0(x)$, the calculation is so involved that we won't consider it in this paper. It is more reasonable to assume a spatial varying chiral condensate in this case, so we can just set $\langle\sigma\rangle=m+m(x)-m_0$. Then by integrating out the fermion degrees of freedom, the partition function can be expressed as a bosonic version:
\begin{eqnarray}
{\cal Z} &=&\int[D\hat{\sigma}][D\hat{\pi}_0][D\hat{\pi}_\pm]\nonumber\\
 &&\!\!\exp\Big\{-\int d^4X\Big[ {\left(m+m(x)-m_0+\hat{\sigma}\right)^2+\hat{\pi}_0^2+\hat{\pi}_\pm^2\over 4G}\Big]\\
&&\!\!+\text {Tr}\ln\left[i{\slashed D}-m-m(x)-\hat{\sigma}-i\gamma^5\left(\tau_3\hat{\pi}_0+\tau_\pm\hat{\pi}_\pm\right)+{\mu}\gamma^0\right]\Big\},\nonumber
\end{eqnarray}
where the fields with hat denote the bosonic fluctuation modes and the trace is taken over the quark spin, flavor, color, and the space-time coordinate spaces. In mean field approximation, the thermodynamic potential can be expressed as
\begin{eqnarray}\label{Omega}
\Omega\!=\!{1\over V_4}\left\{\!\!\int\!\! d^4X{(m\!+\!m(x)\!-\!m_0)^2\over 4G}\!-\!\text {Tr}\ln\left[i{\slashed D}\!-\!m\!-\!m(x)\!+\!{\mu}\gamma^0\right]\right\},
\end{eqnarray}
where the four dimensional volume $V_4=\beta V$ with $\beta=1/T$ the inverse temperature and $V$ the spatial volume of the system. In principal, the gap equation can be obtained by the extremal condition $\delta\Omega/\delta m(x)=0$ as
\begin{eqnarray}\label{gap1}
\int d^3X{m\!+\!m(x)\!-\!m_0\over 2G}-\text {Tr}_3{\cal S}_{A}(x)=0,
\end{eqnarray}
where the fermion propagator in the semi-localized magnetic field is given by ${\cal S}_{A}(x)=-\left[i{\slashed D}-m-m(x)+{\mu}\gamma^0\right]^{-1}$ and the coordinate integrals take over all directions but $x$. The gap equation can be separated into two parts: $x$-independent part which gives the expectation value of $m$ and $x$-dependent part which gives the expectation value of $m(x)$. It is not easy to solve the gap equation for spatially varying $m(x)$ as in the study of inhomogeneous FFLO phases~\cite{Bowers:2002xr,Nickel:2009ke,Cao:2015rea,Cao:2015taa,Cao:2016fby}, let alone that the explicit form of $m(x)$ is unknown. For that sake, we first develop a formalism with only constant $m$ (that is $m(x)=0$) to evaluate the thermodynamic potential in the semi-localized magnetic field and then explore $m(x)$ by using Taylor expansion over small $m(x)$ in the weak magnetic field limit.

\subsection{Thermodynamic potential with constant $m$}\label{potential}

It is usually not easy to solve the Dirac equation exactly in the presence of inhomogeneous magnetic field. But for the chosen semi-localized magnetic field $B(x)=B{\rm sech}^2(x/\lambda)$ with $B$ the magnitude and $\lambda$ the scale, we are able to derive an exact solution~\cite{Cangemi:1995ee,Dunne:1997kw}. Because the magnetic field is well confined in the region $(-3\lambda,3\lambda)$ as shown in Fig.\ref{BJx}, we will first choose $L=6\lambda$ as the system size to study the change of the thermodynamic potential due to the presence of inhomogeneous magnetic field. Or one can understand it in another way, that is, the magnetic field spreads all over the space with the centers at $x_0=6n\lambda, (n\in \mathbb{Z})$, then the situation is just equal to the case with a system size $L=6\lambda$ due to the periodicity of the configuration. After developing the whole formalism, we will extend the study to the system with a fixed size.

For brevity, we will proceed with one color and one flavor first. By following the discussions in Ref.~\cite{Cangemi:1995ee,Dunne:1997kw}, the discrete eigenenergy for a given charge $q_{\rm f}$ in the orthogonal dimensions can be presented as:
\begin{eqnarray}
\epsilon_{ns}(p_2)&=&\Big[p_2^2\!+\!(q_{\rm f}B\lambda)^2\!-\!\lambda^{-2}\Big(n\!+\!{1\over2}\!-\!c_s\Big)^2\nonumber\\
&&\!-(p_2q_{\rm f}B\lambda^2)^2\Big(n\!+\!{1\over2}\!-\!c_s\Big)^{-2}\Big]^{1/2},
\end{eqnarray}
where $c_s=\Big|{1\over2}\!+\!s~q_{\rm f}B\lambda^2\Big|$ with $s=\pm$ denoting the fermion spin along $z$ direction and $n$ is constrained to $0\le n\le N_s$ with $N_s={\rm ceiling}\Big(c_s-{3\over2}\!-\!\sqrt{|p_2q_{\rm f}B\lambda^3|}\Big)$. One should notice that in the constant magnetic field limit $\lambda\rightarrow\infty$, $\epsilon_{ns}(p_2)=[(2n+1-s~{\rm sgn}(q_{\rm f}B))|q_{\rm f}B|]^{1/2}$ and $N_s\rightarrow\infty$ for a fixed $p_2$. Besides, there are also contributions from the continuum spectra as we will illuminate soon. 

In the finite temperature and density case, the contribution of fermion loop to the thermodynamic potential can be given as
\begin{eqnarray}
&&\Omega(m^2,q_{\rm f},B,\lambda)\nonumber\\
&=&\!\!-{1\over2L}\!\int\!\!{d^3p\over(2\pi)^2}{\rm Tr}\ln[\!-\partial^2_x\!+\!V_{p_2}(x)\!+\!m^2\!+\!p_3^2\!+\!(\omega_l\!+\!i\mu)^2]\nonumber\\
&=&\!\!-{1\over2L}\!\int\!\!{d^3p\over(2\pi)^2}\!\!\int\!\!{dm^2} {\rm Tr}[\!-\partial^2_x\!+\!V_{p_2}(x)\!+\!m^2\!+\!p_3^2\!+\!(\omega_l\!+\!i\mu)^2]^{-1}\nonumber\\
&&V_{p_2}(x)=-{\lambda^{-2}}\Big[\Big({1\over2}+q_{\rm f}B\lambda^2\sigma_3\Big)^2-{1\over4}\Big]\Big[1-\tanh^2\Big({x\over \lambda}\Big)\Big]\nonumber\\
&&\ \ \ \ \ \ \ \ \ \ \ \ \ \ \ \ \ +\sum_{t=\pm}{1\over2}(p_2-tq_{\rm f}B\lambda)^2\Big[1+t\tanh\Big({x\over \lambda}\Big)\Big],
\end{eqnarray}
where the fermion Matsubara frequency $\omega_l=(2l+1)\pi T\ (l\in\mathbb{Z})$ and we denote $\int d^3p=\int dp_2dp_3T\sum_{l=-\infty}^{\infty}$ for convenience. Then, the trace of the Green's function can be completed with the help of hypergeometric functions to give
\begin{eqnarray}\label{omega}
&&\Omega(m^2,q_{\rm f},B,\lambda)\nonumber\\
&=&\!\!{\lambda^2\over4L}\int\!{{d^3p\over(2\pi)^2}}\int\!{dm^2} \sum_{s,t=\pm}\Big({1\over\alpha_+}\!+\!{1\over\alpha_-}\Big){\psi\Big({1\over2}(\alpha_+\!+\!\alpha_-\!+\!1)\!-\!t~c_s\Big)}\nonumber\\
&=&{1\over L}\int\!{d^3p\over(2\pi)^2}\int\!{dm^2} \sum_{s,t=\pm}{\partial_{(\omega_l\!+\!i\mu)^2}\Gamma\Big({1\over2}(\alpha_+\!+\!\alpha_-\!+\!1)\!-\!t~c_s\Big)\over \Gamma\Big({1\over2}(\alpha_+\!+\!\alpha_-\!+\!1)\!-\!t~c_s\Big)},
\end{eqnarray}
where $\alpha_\pm={\lambda}\sqrt{E_{\pm}^2(p_2,p_3,m)+(\omega_l+i\mu)^2}$ with the continuum spectrum $E_{t}(p_2,p_3,m)=\sqrt{(p_2+tq_{\rm f}B\lambda)^2+p_3^2+m^2}$. Completing the summation over $\omega_l$ by deforming the integral contour and then the integral over $m^2$, we find the contribution from the bound states or discrete spectra is
\begin{eqnarray}\label{omegab}
&&\Omega_b(m^2,q_{\rm f},B,\lambda,T,\mu)\nonumber\\
&=&-{1\over L}\int{dp_2dp_3\over(2\pi)^2}\int{dm^2} \sum_{s,t=\pm}\sum_{n=0}^{N_s}{\tanh\Big({E_{ns}(p_2,p_3,m)+t\mu\over2T}\Big)\over4E_{ns}(p_2,p_3,m)}\nonumber\\
&=&-{1\over2L}\!\int{dp_2dp_3\over(2\pi)^2}\!\!\sum_{s,t=\pm}\!\sum_{n=0}^{N_s}\Big[E_{ns}\!+\!2T\ln(1\!+\!e^{-(E_{ns}+t\mu)/T})\Big],
\end{eqnarray}
where the dispersion relation is $E_{ns}(p_2,p_3,m)=\Big(\epsilon_{ns}^2(p_2)+p_3^2+m^2\Big)^{1/2}$. If the integral region of $p_2$ is fixed to $\pm q_{\rm f}BL/2$ with $L$ the fixed system size, then in the constant magnetic field limit $\lambda\rightarrow\infty$, we can recover the un-regularized form~\cite{Cao:2015xja} of thermodynamic potential from Eq.(\ref{omegab}). The contribution from the cut branches $\pm E_{t}(p_2,p_3,m)$ can be evaluated with the help of the following properties:
\begin{eqnarray}
&&{1\over2\pi i}\int_{-\infty+i\eta}^{-a+i\eta}dx{f(x^2)\over\sqrt{a^2-x^2}}\tanh\Big({x+\mu\over2T}\Big)\nonumber\\
&=&{1\over2\pi}\int_{a}^\infty dx{f(x^2)\over\sqrt{x^2-a^2}}\tanh\Big({x-\mu\over2T}\Big),\nonumber\\
&&{1\over2\pi i}\int_{a+i\eta}^{\infty+i\eta}dx{f(x^2)\over\sqrt{a^2-x^2}}\tanh\Big({x+\mu\over2T}\Big)\nonumber\\
&=&{1\over2\pi}\int_{a}^\infty dx{f(x^2)\over\sqrt{x^2-a^2}}\tanh\Big({x+\mu\over2T}\Big)
\end{eqnarray}
with $a>0,\eta\gtrapprox0$. Then, the contribution from the continuum spectrum can be given as
\begin{widetext}
\begin{eqnarray}
\Omega_c(m^2,q_{\rm f},B,\lambda,T,\mu)
&=&{\lambda\over16\pi L}\int{dp_2dp_3\over(2\pi)^2}\int{dm^2} \sum_{s,t,u,v=\pm}\int_{E_{t}}^\infty d\omega{1\over\sqrt{\omega^2-E_{t}^2}}\left[\tanh\Big({\omega-\mu\over2T}\Big)+\tanh\Big({\omega+\mu\over2T}\Big)\right]\nonumber\\
&&\psi\left({1\over2}\Big({iv\lambda}\sqrt{\omega^2-E_{t}^2}\!+\!{\lambda}\sqrt{\Big|\omega^2-E_{-t}^2\Big|}\Big(iv~\theta(\omega-{E_{-t}})\!+\!\theta({E_{-t}}-\omega)\Big)\!+\!1\Big)\!-\!u\Big|{1\over2}\!+\!s~q_{\rm f}B\lambda^2\Big|\right)\nonumber\\
&=&{\lambda\over2\pi L}\int{dp_2dp_3\over(2\pi)^2}\sum_{s,u,v=\pm}\int_{0}^\infty {dy}~\left[\sqrt{y^2+E_\bot^2}+T\ln(1+e^{-(\sqrt{y^2+E_\bot^2}+\mu)/T})+T\ln(1+e^{-(\sqrt{y^2+E_\bot^2}-\mu)/T})\right]\nonumber\\
&&\psi\left({1\over2}\Big({iv\lambda}y\!+\!{\lambda}\sqrt{|h|}~\Big(iv~\theta(h)\!+\!\theta(-h)\Big)\!+\!1\Big)\!-\!u\Big({1\over2}\!+\!s~q_{\rm f}B\lambda^2\Big)\right), 
\end{eqnarray}
\end{widetext} 
where $h(y,p_2,q_{\rm f},B,\lambda)=y^2+4p_2q_{\rm f}B\lambda-4(q_{\rm f}B\lambda)^2$ and $E_\bot(p_2,p_3,m)=\sqrt{p_2^2+p_3^2+m^2}$. This actually cannot reduce to the well known form in the vanishing magnetic field limit $B\rightarrow0$ because some $B$ independent terms have been dropped in deriving Eq.(\ref{omega})~\cite{Cangemi:1995ee}. Though, we can still recognize the main part of the thermodynamic potential with eigenenergy $E(p)=\sqrt{p^2+m^2}$ except for the multiplicative digamma function $\psi$.

For further convenience, I denote the vacuum and thermal parts of the thermodynamic potential $\Omega_{b/c}(m^2,q_{\rm f},B,\lambda,T,\mu)$ by $\Omega_{b/c}(m^2,q_{\rm f},B,\lambda)$ and $\Omega_{b/c}^t(m^2,q_{\rm f},B,\lambda,T,\mu)$, separately. The divergence comes solely from the vacuum part $\Omega_B=\Omega_{b}+\Omega_{c}$, the $B$-dependent part of which was renormalized to a compact form by dropping some $B$ independent and $B^2$ terms~\cite{Dunne:1997kw}.  Actually, for the study of chiral symmetry breaking and restoration, the $B^2$ terms can not be dropped at will, otherwise $\Omega_B$ will have a "wrong" sign compared to the case with constant magnetic field~\cite{Cao:2015xja}:
\begin{eqnarray}
\Omega_B={1\over8\pi^2}\int_0^\infty {ds\over s^3}~e^{-m^2s}\left({q_{\rm f}Bs\over\tanh(q_{\rm f}Bs)}-1\right).
\end{eqnarray}
Thus, from the $2+1$ dimensional result~\cite{Cangemi:1995ee}:
\begin{widetext} 
	\begin{eqnarray}\label{omegaB2}
	\Omega_B(m^2,q_{\rm f},B,\lambda)={1\over2\pi\lambda^2L}\int_0^\infty{dx\over e^{2\pi x}-1}\Re\left[\Big(|q_{\rm f}B\lambda^2|-ix\Big)g^{-1}(x)\Big(\lambda^2m^2+g^2(x)\Big)\ln{\lambda m-ig(x)\over \lambda m+ig(x)}\right],
	\end{eqnarray}
we should just keep the third momentum integral form for the $3+1$ dimensional case as~\cite{Dunne:1997kw}:
	\begin{eqnarray}
	\Omega_B(m^2,q_{\rm f},B,\lambda)={1\over2\pi^2\lambda^2L}\int_0^\infty dp_3\int_0^\infty{dx\over e^{2\pi x}-1}\Re\left[\Big(|q_{\rm f}B\lambda^2|-ix\Big)g^{-1}(x)\Big(\lambda^2(m^2+p_3^2)+g^2(x)\Big)\ln{\lambda\sqrt{m^2+p_3^2}-ig(x)\over \lambda \sqrt{m^2+p_3^2}+ig(x)}\right],
	\end{eqnarray}
where $g(x)=x^2+2ix|q_{\rm f}B\lambda^2|$. Then, $\Delta \Omega_B(m^2,q_{\rm f},B,\lambda)-\Delta \Omega_B(m_1^2,q_{\rm f},B,\lambda)$, where $\Delta \Omega$ denotes the difference between the finite magnetic field result and the one in $B\rightarrow0$ limit, can be shown to be convergent for $m_1\neq0$ and negative divergent for $m_1=0$.  Thus, $m\neq0$ is always favored in magnetic field. 
	
Here, it is illuminative to present the thermodynamic potential for $N$ species fermion systems in $2+1$ dimensions because it's renormalizable in large $N$ expansion~\cite{Rosenstein:1990nm}. In chiral limit, by following the renormalization scheme as in Ref.~\cite{Cao:2014uva}, the thermodynamic potential can be presented as
\begin{eqnarray}\label{omega2}
\Omega/N&=&-{m^2m_g\over 2\pi}{\rm sgn}(g-g_c)+{|m|^3\over3\pi}+\Delta\Omega_B,
\end{eqnarray}
where $\Omega_B$ is given by Eq.(\ref{omegaB2}), $m_g$ stands for the magnitude of the coupling $g$ and $g_c$ is the critical coupling constant. One can easily check that $\Delta\Omega_B(m^2,q_{\rm f},B,\lambda)=0$ in the limit $\lambda\rightarrow0$ as should be and the thermodynamic potential is reduced to
\begin{eqnarray}
\Omega(m^2,q_{\rm f},B,\lambda)/N=-{m^2m_g\over 2\pi}{\rm sgn}(g-g_c)\!+\!{m^3\over3\pi}\!+\!{\lambda\over2\pi L}\int_0^\infty\!\!{dx\over x(e^{2\pi x}-1)}\Re\left[\sqrt{|q_{\rm f}B|x}\Big((1\!+\!i)|q_{\rm f}B|x\!+\!{1\!-\!i\over2}m^2\Big)\ln{m\!+\!(1\!-\!i)\sqrt{|q_{\rm f}B|x}\over m\!-\!(1\!-\!i)\sqrt{|q_{\rm f}B|x}}\right],
\end{eqnarray}
in the limit $\lambda\rightarrow\infty$.
\end{widetext}

\begin{figure}[!htb]
	\begin{center}
		\includegraphics[width=8cm]{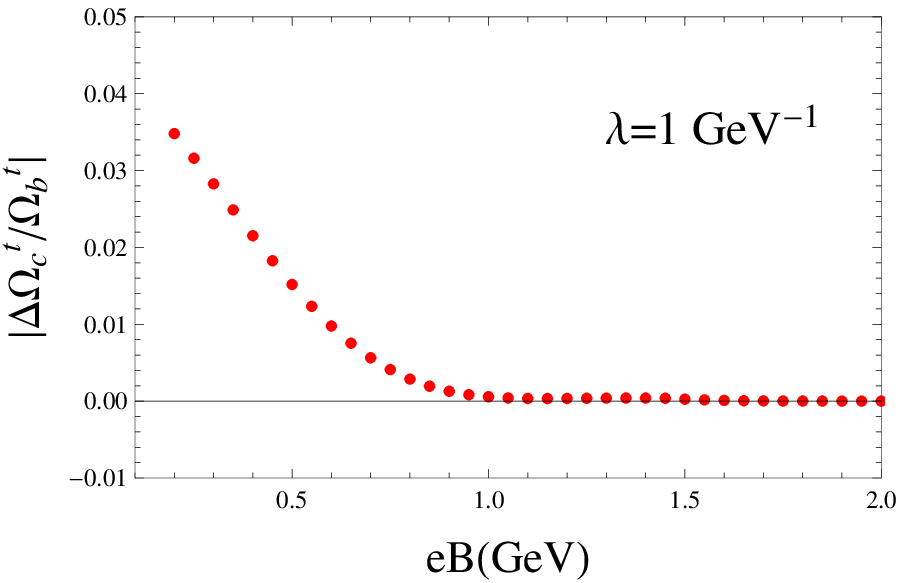}
		\includegraphics[width=7.8cm]{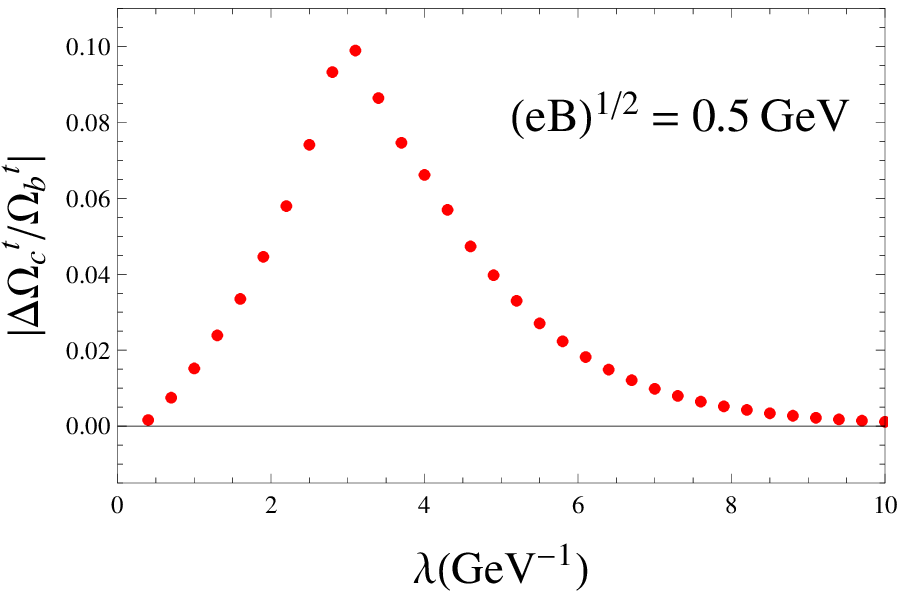}
		\caption{The ratio $\Delta\Omega_c^t/\Omega_b^t$, where the color and flavor degrees of freedom are both taken into account, as a function of the magnetic field magnitude $B$ for a given scale $\lambda$ (upper panel) and of $\lambda$ for a given $B$ (lower panel). The chemical potential, temperature and mass are reasonably chosen as $\mu=0, T=0.15 {\rm GeV}$ and $m=0.3 {\rm GeV}$.\label{ROmega}}
	\end{center}
\end{figure}
Turn back to the case with $3+1$ dimensions. For the thermal parts, it is clear that 
\begin{eqnarray}
\lim_{B\rightarrow0}\Omega_{b}^t(m^2,q_{\rm f},B,\lambda,T,\mu)&=&0,\nonumber\\
\lim_{B\rightarrow0}\Omega_{c}^t(m^2,q_{\rm f},B,\lambda,T,\mu)&\neq&0,
\end{eqnarray}
so the pure $B$-dependent part can be given as
\begin{eqnarray}
\Omega_B^t=\Omega_{b}^t+\Delta\Omega_{c}^t,
\end{eqnarray} 
which vanishes at $B=0$. In order to explore the relative importance of the discrete and continuum eigenstates, I compute the ratio $\Delta\Omega_{c}^t/\Omega_{b}^t$ in this part for convenience. As shown in Fig.\ref{ROmega}, the contribution from continuum part is usually very small compared to the discrete one and can be neglected for simplicity, which justifies the later treatment in the weak magnetic field approximation.
Finally, by recovering the $B$-independent term or the thermodynamic potential in the $B=0$ case, which takes the following three-momentum cutoff regularized form~\cite{Cao:2015dya}
\begin{eqnarray}\label{omegaL}
&&\!\!\!\!\!\!\!\!\!\!\!\!\Omega_\Lambda(m,T,\mu)=-{T\over\pi^2}\sum_{s=\pm}\int_0^\infty p^2 dp \ln\Big(1+e^{-(E(p)+s\mu)/T}\Big)\nonumber\\
&&\!\!\!-{m^3\over8\pi^2}\left[\Lambda\left(1\!+\!{2\Lambda^2\over m^2}\right)\sqrt{1\!+\!{\Lambda^2\over m^2}}\!-\!m\ln\left({\Lambda\over m}
\!+\!\sqrt{1\!+\!{\Lambda^2\over m^2}}\right)\right],
\end{eqnarray}
the total finite thermodynamic potential of NJL model in the given semi-localized magnetic field is
\begin{eqnarray}\label{omega3}
\Omega={(m-m_0)^2\over4G}+N_c\sum_{\rm f=u,d}\Big(\Omega_\Lambda+\Omega_B+\Omega_{B}^t\Big).
\end{eqnarray}

Now, if we fix the system size $L$ which is large enough to neglect boundary effect. Then, only the case $\lambda\gtrsim L$ is interesting for such a system because the average effect of magnetic field is not vanishingly small. In this case, $N_s$ is usually very large, the contribution from the continuum spectrum can be safely neglected due to either the heaviness or the relatively small effective integral region of $p_2$, and the thermodynamic potential is simply given by $\Omega_b(m^2,q_{\rm f},B,\lambda,T,\mu)$ with the integral limit of $p_2$ fixed as $\pm q_{\rm f}BL/2$ as stated before. Then, the thermodynamic potential can be regularized in the more convenient Pauli-Villars scheme~\cite{Cao:2015xja} in this case as:
\begin{eqnarray}
\Omega=\!\!{(m\!-\!m_0)^2\over4G}\!+\!N_c\!\!\sum_{\rm f=u,d}\sum_{j=0}^{2}C_j~\Omega_b(m^2\!+\!j\Lambda^2,B,\lambda,T,\mu)
\end{eqnarray}
with $C_j=3j^2-6j+1$. 

\subsection{Weak magnetic field approximation}\label{WB}
Due to the difficulty in determining the exact mass gap from the gap equation Eq.(\ref{gap1}), I will try to solve this issue in the weak magnetic field limit -- "weak" actually just means the effect of magnetic field is very small compared to the chiral condensate already developed in the vacuum. It is reasonable to expect that the coordinate dependent part of the mass gap $m(x)$ is also restricted to the region where the semi-localized magnetic field exists. So, weak magnetic field just means small $m(x)$ and we can make Taylor expansions of the thermodynamic potential to the second order of $m(x)$, that is,
\begin{eqnarray}
\Omega\!\!&=&\!\!{1\over V_4}\left[\int d^4X{(m+m(x)-m_0)^2\over4G}-{\rm Tr}\ln G^{-1}(X,X)+\right.\nonumber\\
&&\!\!\!\!\left.{\rm Tr}G(X,X)m(x)\!+\!{1\over2}{\rm Tr}G(X,X')m(x')G(X',X)m(x)\right].
\end{eqnarray}
Here, the inverse fermion propagator $G^{-1}(X,X')=i\slashed{D}+\mu\gamma^0-m$ and the constant mass $m$ should be determined in the case without magnetic field with the thermodynamic potential:
\begin{eqnarray}
\Omega={(m-m_0)^2\over4G}-{N_cN_f\over V_4}{\rm tr}\ln\Big(i\slashed{\partial}+\mu\gamma^0-m\Big).\\\nonumber
\\\nonumber
\end{eqnarray} 
From the explicit regularized expression Eq.(\ref{omegaL}), we have the following gap equation~\cite{Cao:2015dya}:
\begin{widetext}
\begin{eqnarray}
{m-m_0\over 2G}&=&{N_c m^2\over\pi^2}\left[\Lambda\sqrt{1+{\Lambda^2\over m^2}}-m\ln\left({\Lambda\over m}
+\sqrt{1+{\Lambda^2\over m^2}}\right)\right]-{N_cm\over\pi^2}\sum_{s=\pm}\int_0^\infty p^2 dp {1\over E(p)}{2\over 1+e^{(E(p)+s\mu)/T}}.
\end{eqnarray}

Then, the extremal condition $\delta\Omega/\delta m(x)=0$ gives the following integral equation:
	\begin{eqnarray}\label{gap3}
	{m+m(x)-m_0\over2G}+\int{d^3p\over (2\pi)^3}~{\rm tr}G(p;x,x)+\int{d^3p\over (2\pi)^3}\int dx'{\rm tr}G(p;x,x')m(x')G(p;x',x)=0,
	\end{eqnarray}
where ${\rm tr}G(p;x,x)$ can be evaluated by following the property
\begin{eqnarray}
{\rm tr}\Big(i\slashed{D}+\mu\gamma^0-m\Big)^{-1}&=&{\rm tr}\Big(i(\omega_l+i\mu)\gamma^0-i\slashed{D}_i-m\Big)\Big[(\omega_l+i\mu)^2-D_i^2+m^2+q_{\rm f}\sigma_{\mu\nu}F^{\mu\nu}\Big]^{-1}\nonumber\\
&=&-m~{\rm tr}\Big[(\omega_l+i\mu)^2-D_i^2+m^2+q_{\rm f}\sigma_{\mu\nu}F^{\mu\nu}\Big]^{-1}
\end{eqnarray}
as~\cite{Cangemi:1995ee,Dunne:1997kw}:
\begin{eqnarray}
&&{\rm tr}G(p;x,x)=-{2m\over W}\sum_{s=\pm} {g_1(s,x)g_2(s,x)},\ W={2\over{\lambda}}{\Gamma(1+2a)\Gamma(1+2b)\over \Gamma(a+b+{1\over2}-c_s)\Gamma(a+b+{1\over2}+c_s)},\nonumber\\
&&g_1(s,x)=\xi^a(1-\xi)^bF(a+b+{1\over2}-c_s,a+b+{1\over2}+c_s;1+2a;\xi),\nonumber\\
&&g_2(s,x)=\xi^a(1-\xi)^bF(a+b+{1\over2}-c_s,a+b+{1\over2}+c_s;1+2b;1-\xi),\ \xi={1+\tanh\Big({x\over\lambda}\Big)\over2}\nonumber\\
&&a={\lambda\over2}\sqrt{(p_2-q_{\rm f}B\lambda)^2+(\omega_l+i\mu)^2+p_3^2+m^2},b={\lambda\over2}\sqrt{(p_2+q_{\rm f}B\lambda)^2+(\omega_l+i\mu)^2+p_3^2+m^2}.
\end{eqnarray}
In the limit $B\rightarrow0$, $F(a+b+{1\over2}-c_s,a+b+{1\over2}+c_s;1+2a;\xi)\rightarrow(1-\xi)^{-2a}$ and the trace reduces to $-2m/\sqrt{p_2^2+(\omega_l+i\mu)^2+p_3^2+m^2}$, which is consistent with the usual one obtained in energy-momentum space but $p_1$ is integrated over first here. In the limit $x\rightarrow\infty$ or $\xi\rightarrow1$, the hypergeometric functions become
\begin{eqnarray}
F(a+b+{1\over2}-c_s,a+b+{1\over2}+c_s;1+2a;\xi)&=&{(1-\xi)^{-2b}\over 2b}{\Gamma(1+2a)\Gamma(1+2b)\over \Gamma(a+b+{1\over2}-c_s)\Gamma(a+b+{1\over2}+c_s)},\nonumber\\
F(a+b+{1\over2}-c_s,a+b+{1\over2}+c_s;1+2b;1-\xi)&=&1.
\end{eqnarray}
Then, ${\rm tr}G(p;x,x)$ becomes magnetic field independent after shifting the integral variable $p_2$ in $b$ which indicates $\lim_{x\rightarrow\infty}m(x)=0$ as expected. For the last term on the left-hand side of the gap equation Eq.(\ref{gap3}), the effective integral region is constrained by $m(x)$ or originally by $B(x)$ to order $\lambda$. Thus, for not too large $\lambda$, it is enough to evaluate this term with the fermion propagator in the absence of magnetic field because that only gives next-to-next-to-next order contribution, and we can just take $m(x')\approx m(x)$ as its leading order contribution of the Taylor expansions around $x$. Finally, the integral equation Eq.(\ref{gap3}) can be reduced to an algebra equation:
\begin{eqnarray}\label{gapx}
m(x)=-\left[{1\over2G}+\int{d^4p\over (2\pi)^4}{\rm tr}\Big(i(\omega_l+i\mu)\gamma^0+p^i\gamma^i-m\Big)^{-2}\right]^{-1}\left[\int{d^3p\over (2\pi)^3}~{\rm tr}G(p;x,x)-(B\rightarrow0)\right].
\end{eqnarray}
The prefactor in the expression of $m(x)$ is actually the propagator of $\sigma$ mode at vanishing energy-momentum and can be given directly as~\cite{Klevansky:1992qe}:
\begin{eqnarray}
\left[{1\over2G}\!+\!\int{d^4p\over (2\pi)^4}{\rm tr}\Big(i(\omega_l+i\mu)\gamma^0+p^i\gamma^i-m\Big)^{-2}\right]^{-1}=\left[{1\over2G}-N_cN_f\int_0^\Lambda{dp\over \pi^2}{p^4\over E^3(p)}\!+\!N_cN_f\sum_{s=\pm}\int_0^\infty{dp\over \pi^2}{p^4\over E^3(p)}{1\over 1+e^{(E(p)+s\mu)/T}}\right]^{-1}.
\end{eqnarray}
In principal, the magnetic field dependent part needs further regularization as the prefactor, but the integral over $p_2$ is automatically constrained for not too large $\lambda$, as will be shown in the following.

According to the properties of hypergeometric function, there is no pole in $g_1(s,x)g_2(s,x)$ for $0<\xi<1$. Thus, the poles of ${\rm tr}G(p;x,x)$ solely come from $\Gamma(a+b+{1\over2}-|c_s|)$ for $a+b+{1\over2}-|c_s|=-n\ (n\in \mathbb{Z})$, which correspond to the discrete spectra. Then, the summation over the Matsubara frequency can be completed to give
\begin{eqnarray}
\int{d^3p\over (2\pi)^3}~{\rm tr}G(p;x_1,x_1)&=&-{2\over\lambda}\sum_{s,t=\pm}\int{d^2p\over (2\pi)^2}\sum_{n=0}^{N_s}{m~\Gamma(2|c_s|-n)\over \Gamma(1+2a_n^s)\Gamma(1+2b_n^s)}{(-1)^n\over n!}{a_n^sb_n^s\over|c_s|-{1\over2}-n}\xi^{2a_n^s}(1-\xi)^{2b_n^s}{\tanh\Big({E_{ns}(p_2,p_3,m)+t\mu\over2T}\Big)\over E_{ns}(p_2,p_3,m)}\nonumber\\
&&F(-n,2|c_s|-n;1+2a_n^s;\xi)F(-n,2|c_s|-n;1+2b_n^s;1-\xi),
\end{eqnarray} 
where the discrete $a_n^s$ and $b_n^s$ are respectively:
\begin{eqnarray}
a_n^s&=&{1\over2}~\Big|\left(n\!+\!{1\over2}\!-\!|c_s|\right)\!-\!(p_2q_{\rm f}B\lambda^3)\left(n\!+\!{1\over2}\!-\!|c_s|\right)^{-1}\Big|,\nonumber\\
b_n^s&=&{1\over2}~\Big|\left(n\!+\!{1\over2}\!-\!|c_s|\right)\!+\!(p_2q_{\rm f}B\lambda^3)\left(n\!+\!{1\over2}\!-\!|c_s|\right)^{-1}\Big|.
\end{eqnarray}
As has been mentioned, the contribution from discrete spectra vanishes automatically at zero magnetic field because $N_s<0$. 
From the non-negativity of $N_s$, the integral limits of $p_2$ are found to be constrained as $\pm (c_s-{3/2})^2/|q_{\rm f}B\lambda^3|$ which play natural momentum cutoffs if $\lambda$ is not so large. The integral over $p_3$ is divergent which can be regularized it by the three momentum cutoff $\Lambda$ for simplicity. Still, there is contribution from the continuum spectra which can be neglected after separating out the $B$-independent part, as indicated in the previous section.
 
As a byproduct, the magnetic field dependent term can be derived directly by neglecting the degree of freedom along $p_3$ in $2+1$ dimensions, that is,
\begin{eqnarray}
\int{d^2p\over (2\pi)^2}~{\rm tr}G(p;x,x)&=&-{2\over\lambda}\sum_{s,t=\pm}\int{dp_2\over 2\pi}\sum_{n=0}^{N_s}{m~\Gamma(2|c_s|-n)\over \Gamma(1+2a_n^s)\Gamma(1+2b_n^s)}{(-1)^n\over n!}{a_n^sb_n^s\over|c_s|-{1\over2}-n}\xi^{2a_n^s}(1-\xi)^{2b_n^s}{\tanh\Big({E_{ns}(p_2,0,m)+t\mu\over2T}\Big)\over E_{ns}(p_2,0,m)}\nonumber\\
&&F(-n,2|c_s|-n;1+2a_n^s;\xi)F(-n,2|c_s|-n;1+2b_n^s;1-\xi).
\end{eqnarray} 
Then, as we've already known the prefactor in the gapped phase at zero temperature~\cite{Rosenstein:1990nm} is
\begin{eqnarray}\label{prefactor}
\left[{1\over 2G}+\int{d^3p\over (2\pi)^3}{\rm tr}\Big(i\omega\gamma^0+p^i\gamma^i-m\Big)^{-2}\right]^{-1}={\pi\over m},
\end{eqnarray}
the mass fluctuation Eq.(\ref{gapx}) is simply reduced to
\begin{eqnarray}
m(x)&=&{2\over\lambda}\sum_{s=\pm}\int{dp_2}\sum_{n=0}^{N_s}{\Gamma(2|c_s|-n)\over \Gamma(1+2a_n^s)\Gamma(1+2b_n^s)}{(-1)^n\over n!}{a_n^sb_n^s\over|c_s|-{1\over2}-n}{\xi^{2a_n^s}(1-\xi)^{2b_n^s}\over E_{ns}(p_2,0,m)}\nonumber\\
&&F(-n,2|c_s|-n;1+2a_n^s;\xi)F(-n,2|c_s|-n;1+2b_n^s;1-\xi).
\end{eqnarray} 
Thus, for second-order transitions such as that induced by coupling tunning, local chiral symmetry breaking with $m(x)\neq0$ will be realized; but for first-order transitions such as that induced by chemical potential, $m(x)\propto m$ due to the invalidity of Eq.(\ref{prefactor}), local chiral symmetry is also restored.
\end{widetext}
\section{Numerical Results}\label{results}
Mainly, I devote this section to exploring the coordinate dependent mass fluctuation $m(x)$ briefly in $2+1$ dimensions and in detail in $3+1$ dimensions. In $2+1$ dimensions, there is only one energy scale, $m_g$, in the vacuum, so I take $m_g$ as the unit of all other dimensional quantities for universality. In $3+1$ dimensions, the parameters of the NJL model were fixed to $G=4.93~{\rm GeV}^{-2}$, $\Lambda=0.653~{\rm GeV}$ and $m_0=5~{\rm MeV}$ by fitting the pion mass $m_\pi=134~{\rm MeV}$, pion decay constant $f_\pi=93~{\rm MeV}$ and quark condensate $\langle\bar\psi\psi\rangle=-2\times (0.25~{\rm GeV})^3$ in the vacuum~\cite{Zhuang:1994dw}. 

\begin{figure}[!htb]
	\begin{center}
		\includegraphics[width=8cm]{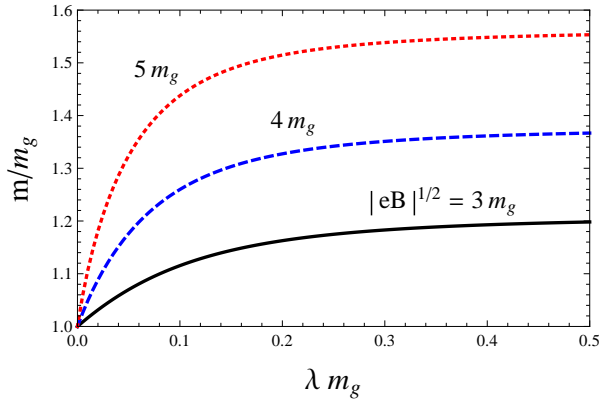}
		\caption{The mass $m$ as a function of the magnetic field scale $\lambda$ for fixed magnitude $B$ at zero temperature in $2+1$ dimensions with supercritical coupling $g>g_c$. All the quantities are scaled by $m_g$ to dimensionless ones.\label{m2D}}
	\end{center}
\end{figure}
\begin{figure}[!htb]
	\begin{center}
		\includegraphics[width=8cm]{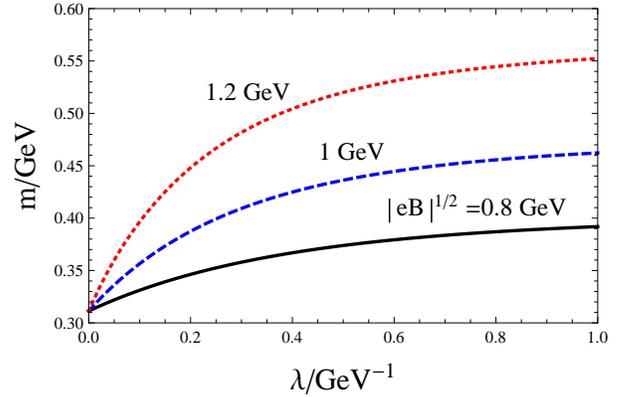}
		\caption{The mass $m$ as a function of the magnetic field scale $\lambda$ for fixed magnitude $B$ at zero temperature in $3+1$ dimensional NJL model.\label{m3D}}
	\end{center}
\end{figure}
Before that, it is instructive to qualitatively illuminate the effects of the magnitude $B$ and scale $\lambda$ of the magnetic field to chiral symmetry breaking and restoration in constant $m$ ansatz. The gap equations can be derived from the thermodynamic potentials Eq.(\ref{omega2}) and Eq.(\ref{omega3}) through $\partial \Omega/\partial m=0$ for $2+1$ and $3+1$ dimensions, respectively. The results are shown in Fig.\ref{m2D} and Fig.\ref{m3D}, from which both chiral catalysis effects of $B$ and $\lambda$ can be easily identified.

\begin{figure}[!htb]
	\begin{center}
		\includegraphics[width=8.2cm]{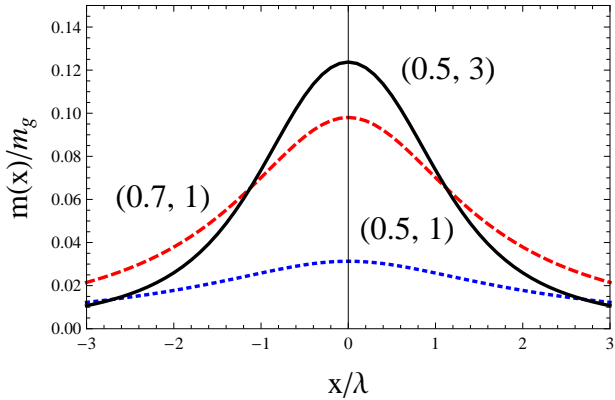}
		\caption{The mass gap $m(x)$ in the region where the magnetic field is localized in $2+1$ dimensions at zero temperature and baryon chemical potential. The parameters shown in the plot are $(|eB|^{1/2}/m_g,\lambda m_g)$.}\label{mx2}
	\end{center}
\end{figure}

Then, for more reasonable study of local chiral symmetry breaking and restoration, the $2+1$ dimensional results are illuminated in Fig.\ref{mx2} for the supercritical case $g>g_c$. As we can see, the weak magnetic field approximation is still good for the magnetic field comparable to $m_g$ and magnetic catalysis effect shows up for the local chiral symmetry breaking. Besides, larger magnetic field scale $\lambda$ usually means higher peak but smaller half-width of $m(x)$. It can be understood in this way: For larger $\lambda$, the region near the original is more like in a constant magnetic field, which of course prefers a larger $m(0)$ due to MCE. All the features are qualitatively consistent with the previous results obtained in constant $m$ ansatz (Fig.\ref{m2D}).

\begin{figure}[!htb]
	\begin{center}
		\includegraphics[width=8.2cm]{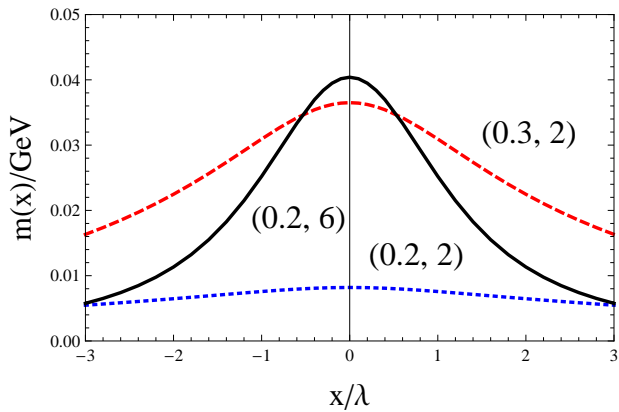}
		\caption{The mass gap $m(x)$ in the region where the magnetic field is localized in $3+1$ dimensional NJL model at zero temperature and baryon chemical potential. The parameters shown in the plot are $(|eB|^{1/2},\lambda)$ in the unit $({\rm GeV},{\rm GeV}^{-1})$.}\label{mx3}
	\end{center}
\end{figure}
\begin{figure}[!htb]
	\begin{center}
		\includegraphics[width=8.2cm]{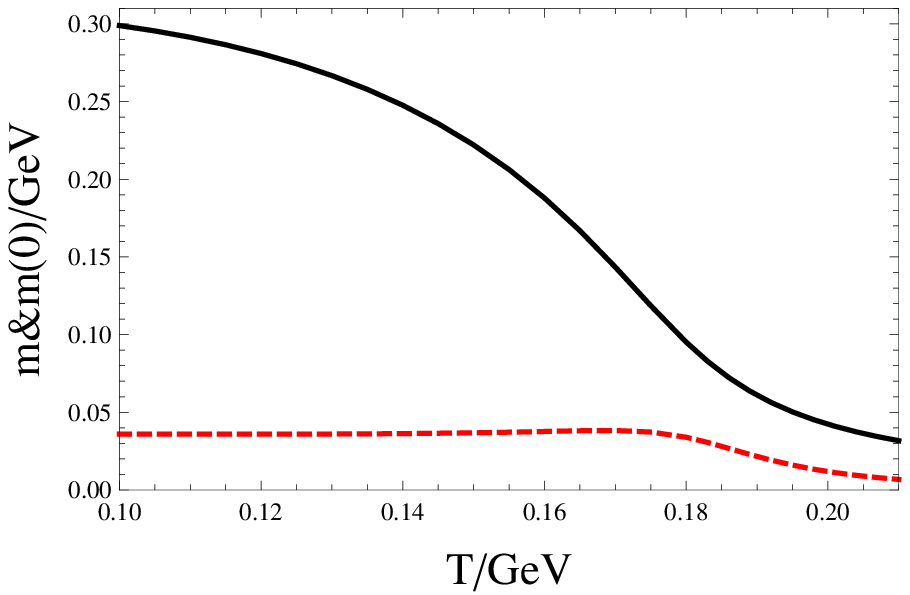}
		\includegraphics[width=8.4cm]{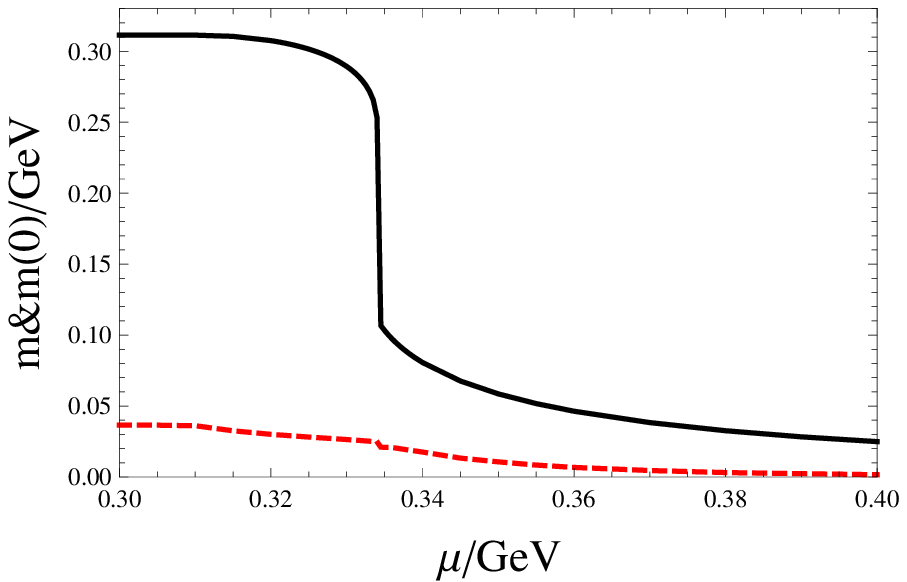}
		\caption{(color online) The constant mass gap $m$ (black solid lines) and the maximal fluctuation $m(0)$ (red dashed lines) as functions of temperature $T$ at vanishing baryon chemical potential (upper panel) and baryon chemical potential $\mu$ at vanishing temperature in $3+1$ dimensional NJL model. The magnitude and scale of the magnetic field are chosen as $(|eB|^{1/2},\lambda)=(0.3{\rm GeV},2 {\rm GeV}^{-1})$.}\label{mTmu}
	\end{center}
\end{figure}

The $3+1$ dimensional results are illuminated in Fig.\ref{mx3} which share similar features as those of the $2+1$ dimensional case and are qualitatively consistent with the results obtained in constant $m$ ansatz (Fig.\ref{m3D}). Then, in order to explicitly show how the local mass fluctuation responds to the global chiral symmetry restoration, we calculate the constant mass $m$ together with the original mass fluctuation $m(0)$ at different temperature and chemical potential as shown in Fig.\ref{mTmu}. As is illuminated, the fluctuation is not sensitive to the change of $m$ when it is still considerably large and is deeply suppressed when it becomes small, which justifies the Taylor expansions in the whole regions. Because of the approximate chiral symmetry with finite current quark mass and non-renormalizability with four fermion couplings of the NJL model, the expected features across the phase transition for $2+1$ dimensions are not found. Specially, the vanishing of $m(x)$ is not found across the first-order transition because $m$ is still finite after the transition.

\section{Conclusions}\label{conclusions}
In this work, I first developed a formalism to evaluate the thermodynamic potential with constant fermion mass in the region where magnetic field is localized and also for the system with fixed size under the framework of NJL model. Then apart from the magnetic field independent terms, the contributions from the discrete and continuum spectra are compared with each other, which indicates the negligible of the latter. Finally, we tried to study the local chiral symmetry breaking due to weak semi-localized magnetic field, which is the main motivation of this work, by using Taylor expansion technique. 

The main findings are the followings. In the constant $m$ ansatz, both the magnetic field magnitude $B$ and scale $\lambda$ tend to catalyze chiral symmetry breaking in $2+1$ and $3+1$ dimensional cases. And in the weak magnetic field approximation, the local chiral symmetry breaking is also found to be enhanced by both $B$ and $\lambda$ which confirms the qualitative features from the constant $m$ ansatz. Thus, the results indicate the importance of inhomogeneous magnetic field effect in HICs with much larger $B$ and that the expanding of fireball ($\lambda$ becomes larger) doesn't necessarily reduce the magnetic field effect during the period when $B$ sustains. Furthermore, the mass fluctuation $m(x)$ is found to be not sensitive to the change of temperature $T$ or baryon chemical potential $\mu$ when the global mass is still considerably large, which further supports the importance of the inhomogeneous magnetic effect in HICs.

\emph{Acknowledgments}---
I thank Xu-guang Huang from Fudan University for his comments on this work. GC is supported by the Thousand Young Talents Program of China, Shanghai Natural Science Foundation with Grant No. 14ZR1403000, NSFC with Grant No. 11535012 and  No. 11675041, and China Postdoctoral Science Foundation with Grant No. KLH1512072.

\end{document}